\documentclass[]{aa}
\usepackage{graphicx}
\usepackage{natbib}
\bibpunct{(}{)}{;}{z}{}{,}
\def\FIR{\ifmmode {\,$\tau_{\rm FIR}$} \else $\,\tau_{\rm FIR}$\fi}
\def\tk{\ifmmode {\,$T_{\rm k}$} \else $\,T_{\rm k}$\fi}
\def\mic{\ifmmode {\,\mu{\rm m}} \else $\,\mu{\rm m}$\fi}       
\def\kms{\ifmmode {\,{\rm km\,s^{-1}}}                          
 \else {\hbox{$\,$ {\rm km$\,$s$^{\rm -1}$}}}\fi}
\def\mo {\ifmmode {\,{\it M}\odot} \else $\,M$\odot\fi} 
\def\lo {\ifmmode {\,{\it L}\odot} \else $\,L$\odot\fi} 
\def\my {\ifmmode {\,{\it M}\solar\,{\rm yr^{-1}}}              
 \else {$\,M$\solar$\,$yr$^{\rm -1}$}\fi}
\def\cmm#1{\ifmmode {\,{\rm cm^{-#1}}}          
 \else \hbox{$\,${\rm cm$^{\rm -#1}$}}\fi}
\chardef\isp="10\def\i{\'\isp}  
\def\as {\ifmmode {^{\scriptscriptstyle\prime\prime}}           
 \else $^{\scriptscriptstyle\prime\prime}$\fi}
\def\am {\ifmmode {^{\scriptscriptstyle\prime}}                 
 \else $^{\scriptscriptstyle\prime}$\fi}
\def\deg {\ifmmode^\circ\else$^\circ$\fi}                       
\def\raw {\ifmmode\rightarrow\else$\rightarrow$\fi}             
\def\x {\ifmmode\times\else$\times$\fi}                         
\def\gsim {\ifmmode {\buildrel>\over\sim}               
 \else {\lower.6ex\hbox{$\buildrel>\over\sim$}}\fi}
\def\lsim {\ifmmode {\buildrel<\over\sim}               
 \else {\lower.6ex\hbox{$\buildrel<\over\sim$}}\fi}
\def\ra[#1 #2 #3.#4]{ #1$^{\rm h}$#2$^{\rm m}$#3$^{\rm s}$.#4}  
\def\dec[#1 #2 #3.#4]{ #1\deg#2\am#3{\as}.#4}             
\def\rax[#1 #2 #3]{RA: #1$^{\rm h}$#2$^{\rm m}$#3$^{\rm s}$}
\def\decx[#1 #2 #3]{Dec:#1\deg#2\am#3\as}          
\def\h2{\rm H$_2$}                      
\begin{document}
\title{Detection of a hot core in the intermediate-mass Class 0 protostar NGC 7129--FIRS 2}
  \author{A. Fuente\inst{1}
\and
  R. Neri\inst{2}
  \and
  P. Caselli\inst{3} 
}
\institute{Observatorio Astron\'omico Nacional (IGN), Campus
 Universitario, Apdo. 112, E-28803 Alcal\'a de Henares (Madrid), Spain     
\and
Institut de Radioastronomie Millim\'etrique, 300 rue de la Piscine, 38406 St. Martin d'H\`eres Cedex, France
\and
INAF-Osservatorio Astrofisico de Arcetri, Largo Enrico Fermi 5, I-50125 Firenze, Italy
} 

\offprints{A. Fuente, \email{a.fuente@oan.es}}
       
%
%
\abstract{
We report high angular resolution (HPBW$\sim$ 0.6$''$ $\times$ 0.5$''$ at 1.3mm) observations 
of the Class 0 intermediate-mass (IM) protostar NGC~7129--FIRS~2 using the Plateau de Bure 
Interferometer. Our observations show the existence of an intense unresolved source in the continuum
at 1.3mm and 3mm at the position of the Class 0 object. In addition, compact CH$_3$CN emission is
detected at this position. The high rotational temperature derived from the CH$_3$CN lines (T$_{rot}$$\approx$50~K)
as well as the enhanced CH$_3$CN fractional abundance (X(CH$_3$CN)$\sim$7.0~10$^{-9}$)
show the existence of a hot core in this
IM young stellar object. This is, up to our knowledge, the first IM hot core detected so far.
Interferometric maps of the region in the CH$_3$OH 5$_{kk'}$$\rightarrow$4$_{kk'}$, 
and D$_2$CO 4$_{04}$$\rightarrow$3$_{03}$ lines are also
presented in this paper. The methanol
emission presents two condensations, one associated with the hot core which is very intense in the
high upper state energy lines (E$_u$$>$100~K) and other associated with the
bipolar outflow which dominates the emission in the low excitation lines. Enhanced CH$_3$OH abundances
(X(CH$_3$OH) $\sim$ 3 10$^{-8}$-- a few 10$^{-7}$) are measured in both components. 
While intense D$_2$CO $4_{04}$$\rightarrow$3$_{03}$ emission is detected towards the hot core, 
the N$_2$D$^+$ 3$\rightarrow$2 line has not been detected in our interferometric observations. 
The different behaviors
of D$_2$CO and N$_2$D$^+$ emissions suggest different 
formation mechanisms for the two species and different deuteration
processes for H$_2$CO and N$_2$H$^+$ (surface and gas-phase chemistry, 
respectively). 
Finally, the spectrum of the large bandwidth correlator show a forest of lines at the hot core position
reavealing that this object is extraordinarily rich
in complex molecules. To have a deeper insight into the chemistry of complex molecules, 
we have compared the fractional abundances of the complex O- and N- bearing species
in FIRS~2 with those in hot corinos and massive hot cores. Within the large uncertainty
involved in fractional abundance estimates towards hot cores, we do not detect any variation of
the relative abundances of O- and N-bearing molecules ([CH$_3$CN]/[CH$_3$OH]) 
with the hot core luminosity. However, 
the O-bearing species H$_2$CO and HCOOH seem to be more abundant in low and intermediate mass
stars than in massive star forming regions. We propose that this could be the consequence of a different grain
mantle composition in low and massive star forming regions.

}
\maketitle

\section{Introduction}
 
Hot cores are compact objects near or around protostars characterized by warm temperatures (T$_k$$>$100 K) and
high densities (n$>$10$^6$~cm$^{-3}$). These regions are also characterized by a very rich chemistry in complex 
molecules (CH$_3$OH, CH$_3$CN, CH$_3$OCHO, CH$_3$OCH$_3$,C$_2$H$_5$CN ...). 
Hot cores are thought to be associated with high-mass protostars (M$\geq$8~M$_\odot$) and 
to represent an important phase in their evolution toward ultracompact and compact HII regions. Recently, 
regions characterized by warm temperatures and high densities have also been detected in 
two low-mass protostars IRAS~16293--2422  \citep{cec00,caz03} and 
NGC~1333~IRAS~4A \citep{bot04}. Complex molecules typical of hot cores 
(e.g. HCOOH, CH$_3$OCHO, CH$_3$CN,
C$_2$H$_5$CN) have also been detected in these objects. 
However, the amount of warm material involved, as well as the chemistry, are different in the two classes of objects.
For this reason, the warm regions in the inner envelope of low mass protostars are usually referred to as $``$hot corinos". 

The formation of complex molecules in hot cores and corinos is poorly 
understood. In the standard scheme, neutral molecules (CO, CS..) are 
frozen onto dust grains during the cold pre-stellar phase. If the dust
temperature is sufficiently low during this phase, surface hydrogenation
of CO leads to the formation of solid H$_2$CO and CH$_3$OH (e.g. Brown et al. 
1988; Charnley et al. 1992; Caselli et al. 1993).  Once the star 
starts heating the grain surfaces, these molecules (called ``parent'' species)
evaporate, enlarging their abundances in the gas phase. 
Because of the high temperature and density of hot cores and corinos, 
these molecules undergo fast neutral-neutral and ion-neutral reactions producing
at early stages ($\sim 10^4$ yr) a second generation of complex O--bearing 
species called $``$daughter" molecules, such as methyl formate, CH$_3$OCHO.  
However,  Horn et al. (2004) recently found that the gas--phase reaction sequence 
between protonated methanol and formaldehyde, crucial for the gas--phase 
formation of methyl formate,  does not proceed in their laboratory 
experiments.  This suggests that surface chemistry is probably playing a key 
role also in the formation of this species. 

Regarding complex N--bearing species (e.g. CH$_3$CN, CH$_3$CH$_2$CN),
they are observed in hot cores and hot corinos. In the chemical scheme of Rodgers 
\& Charnley (2001), complex N--bearing species are thought to be formed 
in the gas phase about 10$^5$ yr after the grain mantle
evaporation, only if the gas temperature is sufficiently large ($\ge$ 300 K).  
On the other hand, Caselli et al. (1993) found that large abundances 
of ethyl cyanide ($X$(CH$_3$CH$_2$CN) $\sim 10^{-7}$, w.r.t. H$_2$) can be 
obtained on grain surfaces, if the dust  temperature during the protostellar
accreting is around 40 K.  In the case of lower temperatures (10--20 K), 
the CH$_3$CH$_2$CN can still be formed on the surface, but with 
reduced abundances ($\sim 10^{-9}$). Thus, Caselli et al.
(1993) suggests that saturated (or H--rich) complex N--bearing species (such as 
ethyl cyanide) are ``parent'' species and that they should be observable 
as soon as the grain mantles are released in the gas phase upon stellar 
heating of the dust. Later on, ``H--poor''  N--bearing molecules (e.g. vinyl 
cyanide, CH$_2$CHCN) can be formed in the gas phase from the destruction of 
CH$_3$CH$_2$CN.  

Indeed, if complex species in general are not formed on grain surfaces, there are 
important problems for understanding the chemical composition of hot corinos: 
the timescale necessary to convert $``$parent" molecules into complex 
$``$daughter" molecules is much longer than the transit time of the gas
in hot corinos (a few hundred years). Therefore, it seems that the chemistry 
of complex molecules must begin on grain surfaces (at least in hot corinos).
However, chemists are still far from a unique interpretation of the hot 
core (and corino) chemistry. More observational work is needed.  In particular, 
given the large sensitivity of surface and gas--phase chemistry on dust  
temperature and gas density, it will be extremely important to measure
possible variations of complex molecule abundances with the physical 
characteristics of hot cores, which of course depend on the mass and 
luminosity of the associated protostar. 

In this paper, we present an interferometric study of the  hot core 
associated with the  
Class 0 IM protostar NGC~7129--FIRS~2. With a luminosity $\sim$ 500 L$_\odot$ and 
a mass $\sim$5~M$_{\odot}$ , FIRS~2  is very likely the
youngest IM object known at present (Fuente et al. 1998,2002). An energetic bipolar molecular outflow with a
quadrupolar morphology is associated with it (Fuente et al. 2001, hereafter Paper I).
Interferometric observations in the continuum at 1mm and the $^{12}$CO 
2$\rightarrow$1 line shows that the quadrupolar morphology of the outflow is due to the 
superposition of two bipolar molecular outflows FIRS~2-out~1 and FIRS~2-out~2.
FIRS~2-out~1 is associated with 
the Class 0 protostar detected as an intense millimeter clump in the
continuum image (MM1 in the nomenclature of Paper I)
while FIRS~2-out~2 is associated with a more evolved infrared star
(FIRS~2~-~IR) undetected at millimeter wavelengths. 

A quite complete chemical study of this Class 0 IM source has been carried out by
Fuente et al. (2005) (hereafter Paper II) using the 30m IRAM telescope. 
They detected warm CH$_3$CN ( T$_k$$>$63~K) towards
the central position which constitutes a strong evidence for the existence
of a hot core in this Class 0 IM object. However, the limited angular resolution
of these observations make it very difficult to discern between  the 
hot core emission and those of the outflow and/or the warm envelope. 
The new results presented in
this paper confirm the existence of the hot core associated with FIRS~2  and 
give a first glance of the chemistry of this object. Up to our knowledge,
FIRS~2 is the first IM hot core detected thus far.
The intermediate kinetic temperature
and mass of the FIRS~2 hot core are expected to produce a differentiated
chemistry and to furnish a link between the low mass and high mass
regimes. 

\section{Observations}

The observations presented in this paper correspond to two different
observational projects carried out with the Plateau de Bure 
interferometer (PdBI).
The main set of data was observed on March 08, 2003.
The observations were carried out with excellent weather conditions 
and counts 5 hr of integration time (on-source) with
6-element array in the A configuration which provides
the highest angular resolution. 
The spectral correlator setup was adjusted to observe
the CH$_3$CN 5$\rightarrow$4 line with two contiguous 20 MHz units, 
N$_2$D$^+$ 3$\rightarrow$2 and D$_2$CO 4$_{04}$$\rightarrow$3$_{03}$
with two separated
20 MHz units and  the remaining units were configured for maximum
continuum sensitivity (see Table 1). These observations show on average system
temperatures of 140 K (3mm) and 250 K (1mm), a mean atmospheric water
vapour content of 1-2mm, and were made under seeing 
conditions of 0.3 arcsec. 
The RF calibrator was 0420-0414, the phase calibrators ordered by
right ascension were 1928+738, 2146+608 and 2309+454, and the flux
calibrator was MWC 349. The precision in the flux densities is better than
20\% at 1mm and better than 10\% at 3mm. Images have been created in natural
weighting. 

In addition, we present the CH$_3$OH J=5$_k$$\rightarrow$4$_k$ image observed
within a previous project in November and
December, 1998. The data corresponding to this project were partially 
published in Paper I. Details
about the observations are given in Table 1 and  Paper I.
  
\setlength\unitlength{1cm}
\begin{figure}
\vspace{12cm}
\includegraphics{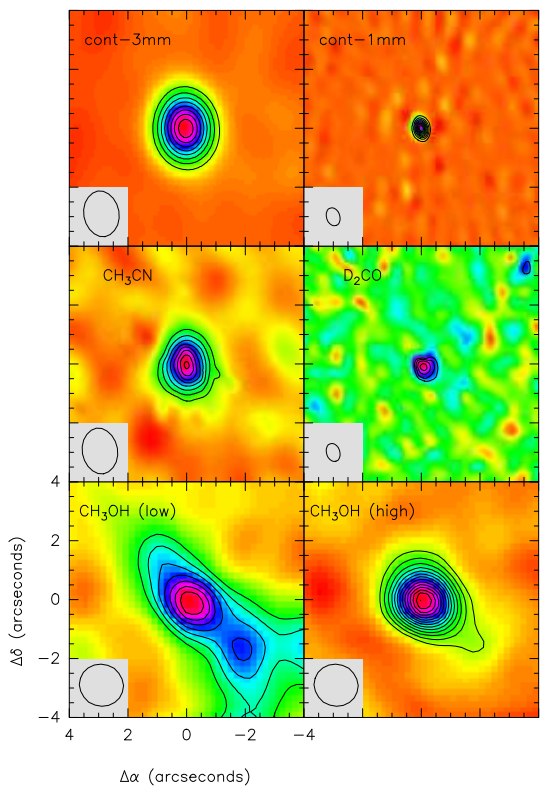}
\caption{Interferometric maps of the Class 0 protostar NGC~7129--FIRS~2 in the
continuum at 1mm and 3mm, and the following molecular lines: CH$_3$CN 5$\rightarrow$4, 
D$_2$CO 4$_{04}$$\rightarrow$3$_{03}$, CH$_3$OH 
5$_{1,4}$$\rightarrow$4$_{1,4}$. The panel marked with $``$CH$_3$OH (high)"
show the integrated intensity map of the lines 5$_{3,1}$$\rightarrow$4$_{3,1}$,
5$_{3,2}$$\rightarrow$4$_{3,2}$,5$_{2,2}$$\rightarrow$4$_{2,2}$
and 5$_{3,3}$$\rightarrow$4$_{3,3}$. Contour levels are: a) 10 to 60 by 5 mJy/beam;
b) 20 to 160 by 20 mJy/beam ;c) 0.6 to 2.7 by 0.3 Jy/beam~km~s$^{-1}$; 
d) 0.3 to 0.6 by 0.1 Jy/beam~km~s$^{-1}$;
e) 1 to 4 by 0.5 Jy/beam~km~s$^{-1}$;
f) 1 to 7.5 by 0.5 Jy/beam km s$^{-1}$.}
\label{fig:maps}
\end{figure}
  
\setlength\unitlength{1cm}
\begin{figure}
\vspace{12.5cm}
\includegraphics{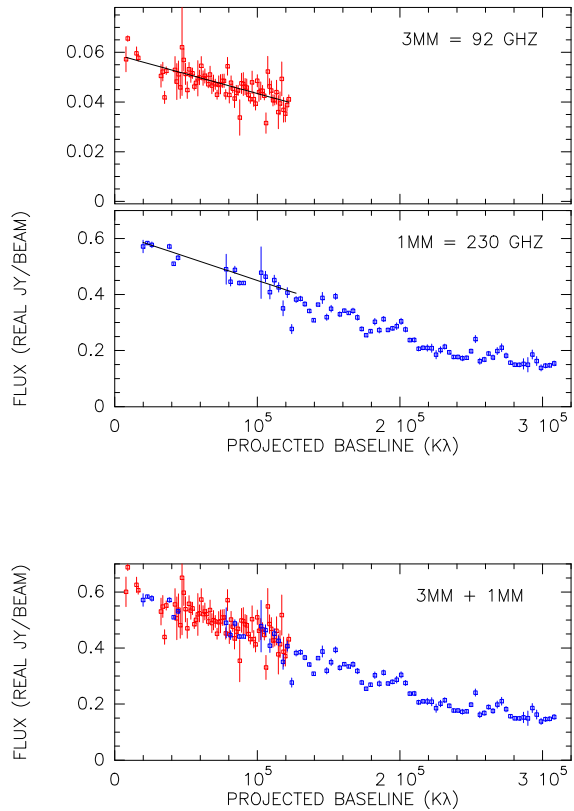}
\caption{Continuum visibilities at 92 GHz and 230 GHz vs projected baselines
in units of wavelength. In the bottom panel we show together the 92 GHz 
and  the 230 GHz visibilities. The 92 GHz visibilities have been scaled by a 
factor of 10.5 to match the 230 GHz ones. The perfect match between the
92 GHz and 230 GHz visibilities suggests that both emissions arise in the
same region as expected in the case of 
dust thermal emission. }
\end{figure}

\section{Continuum data}

The continuum images at 3mm and 1.3mm are shown in Fig. 1. 
The angular resolution of these images is  3 times better than that 
of the images previously published in Paper~I. 
Both images show an intense and compact
source located at RA(2000)=21:43:01.684, Dec(2000)=66:03:23.619,
the position named MM1 in Paper I. There is no indication
of the presence of the second and more extended
source MM2 which is very likely resolved out in the high angular 
resolution image. 

We have modeled the
visibilities to have a deeper insight into the source structure.   
In Fig. 2 we show the 92~GHz and 230~GHz visibilities vs projected baselines
in units of wavelength. The 92 GHz visibilities scaled by a factor of
10.5 match perfectly the 230 GHz visibilities. This suggests that both
emissions are arising exactly in the same region. The spectral index is
$\alpha$=2.56. The fact that the 3mm and 1.3mm emission arises in the
same region and the spectral index larger than 2 is consistent with dust thermal
emission. We have used the high-angular resolution 1.3mm image to model the
continuum emission. Several models (elliptical Gaussian, disk, F$_\nu$=$\nu^{-2}$,
F$_\nu$=$\nu^{-3}$, and an elliptical Gaussian+point source) have been
used to fit the visibilities. The best fit has been obtained with the
elliptical Gaussian+point source model. The HPFW of the elliptical Gaussian is
$\sim$650$\times$900~AU and the flux is 0.43~Jy. The point source has a flux of 0.13~Jy.
The total flux of the compact 1.3~mm component is $\sim$0.56~Jy. This implies a total
(dust +gas) mass of 2~M$_\odot$ assuming a standard dust temperature of 100~K and a dust
emissivity $k_\nu$=0.015(1300/$\lambda$ ($\mu$m)) cm$^2$~g$^{-1}$. Taking into 
account the uncertainty in the value of the dust temperature which could
range between 50~K and 300~K this mass is accurate within a factor of 4. 

We speculate about the possibility that the point source is an accretion disk. In this case,
the mass of the disk would be $\sim$0.3--0.8~M$_\odot$. This value is 30 times larger
than that found by Fuente et al. (2003) in the Herbig Be star R Mon, but it is similar to
that found by Rodr{\i}guez et al. (2005) in the Class 0 low-mass protostar IRAS~16293--2422B. 
Thus, if our assumption is confirmed,
the difference in the disk occurrence and masses between IM and low mass stars
would be related to a short timescale for the disk dissipation 
instead of to 
differences in the first stages of the star formation. This interpretation
is in line with  Fuente et al. (2003).
 
{\scriptsize
\begin{table}[]
\caption{Observations with the Plateau de Bure Interferometer (PdBI)}
\begin{tabular}{l ccc}
\\ \hline
\multicolumn{1}{c}{} & 
\multicolumn{1}{c}{Conf.} &       
\multicolumn{1}{c}{Beam} &
\multicolumn{1}{c}{date}  \\
\hline
Continuum 3mm & A & 1.56$''$$\times$1.20$''$ &  March 03  \\
Continuum 1mm & A & 0.63$''$$\times$0.46$''$ &  March 03 \\
CH$_3$CN 5$_k$$\rightarrow$4$_k$  &  A  & 1.56$''$$\times$1.20$''$ & March 03   \\
N$_2$D$^+$ 3$\rightarrow$2 &  A & 0.63$''$$\times$0.46$''$ &  March 03 \\
D$_2$CO 4$_{04}$$\rightarrow$3$_{03}$  & A &  0.63$''$$\times$0.46$''$ & March 03 \\
CH$_3$OH 5$_{k,k'}$$\rightarrow$4$_{k,k'}$  &  CD & 1.51$''$$\times$1.42$''$ &   Nov, Dec 98 \\   \hline
\end{tabular}
\end{table}
}
    
{\scriptsize
\begin{table*}[]
\caption{Gaussian fits to the line and continuum emission}
\begin{tabular}{l cccc}
\\ \hline
\multicolumn{1}{c}{} & 
\multicolumn{2}{c}{Position$^a$} &       
\multicolumn{1}{c}{Integrated intensity$^b$} &
\multicolumn{1}{c}{Size$^c$}  \\
\multicolumn{1}{c}{} & 
\multicolumn{1}{c}{} &       
\multicolumn{1}{c}{} &       
\multicolumn{1}{c}{(Jy)} &
\multicolumn{1}{c}{(arcseconds)}  \\
\hline
Continuum 3mm & 21:43:01.7 & 66:03:23.6 &  0.0565(0.0005) & 0.66(0.02)$\times$0.54(0.02) \\ 
Continuum 1mm & 21:43:01.7 & 66:03:23.7 &  0.4278(0.0047) & 0.72(0.01)$\times$0.52(0.01) \\
                            & 21:43:01.7 & 66:03:23.6 &  0.1276(0.0046) & Point source ($<0.3"$) \\
CH$_3$CN 5$_k$$\rightarrow$4$_k$  &  21:43:01.7 & 66:03:23.6 & 0.0734(0.0060)  & 0.63(0.22)$\times$0.46(0.19)   \\
D$_2$CO 4$_{04}$$\rightarrow$3$_{0,3}$ & 21:43:01.7 & 66:03:23.6 & 0.76(0.6) & 0.58(0.24)$\times$0.40(0.24) \\
\hline
\end{tabular}

\noindent
$^a$ The absolute positional precision is $\le 0.3"$.

\noindent
$^b$ Systematic calibration errors are not included in the error budget associated to the
integrated intensity 

\noindent
$^c$ Deconvolved by the synthesized interferometric beam.
\end{table*}
}

\setlength\unitlength{1cm}
\begin{figure*}
\vspace{14cm}
\includegraphics{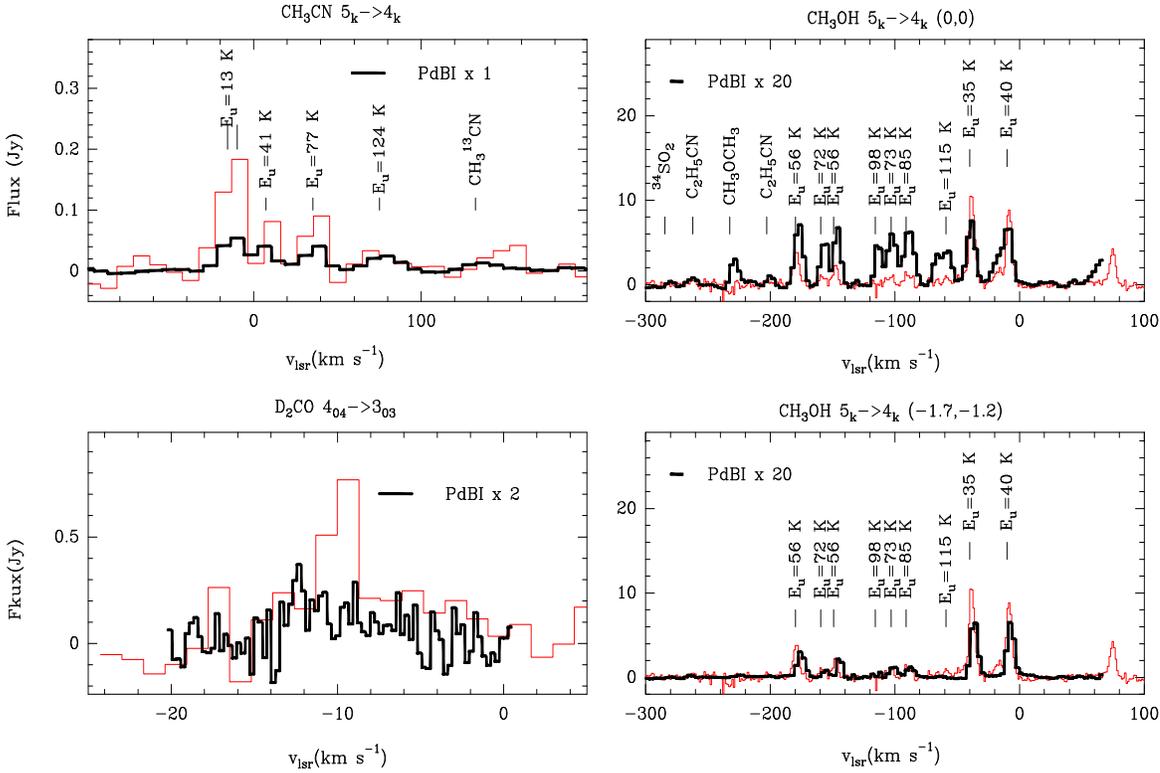}
\caption{Comparison between the single-dish (thin histograms or red) 
and interferometric (thick histograms or black) spectra towards
the hot core position. The interferometric spectrum has been scaled for an easier 
comparison. Note that we have recovered all the flux for the CH$_3$CN line with
the highest upper state energy.}
\end{figure*}
 
\section{Molecular line images}
\subsection{CH$_3$CN}
The interferometric image of the Class 0 protostar NGC~7129--FIRS~2
in the CH$_3$CN 5$\rightarrow$4 line shows a compact source
at the same position and with the same size as the
1mm continuum source (Fig.~1). 
The total agreement in position and size between the continuum and
the CH$_3$CN emissions points out to the existence of a well-differentiated
compact component in the protostellar envelope. 

Because of the rotational structure of CH$_3$CN, one can observe several 
lines at different energies very close in frequency. In Fig. 3 we show the interferometric spectrum
of the CH$_3$CN 5$\rightarrow$4 line compared to the single-dish spectrum reported
in Paper II.  The interferometer has recovered 100\% of the flux of the
highest upper state energy component (E$_u$=124~K) (see Fig. 3). 
However, only 37\% of the flux of the lowest energy
component (E$_u$=13~K) has been recovered (see Fig. 3). 
This suggests that while the emission of the low energy CH$_3$CN lines 
mainly arises in the cooler and more extended envelope, the emission of the high 
energy lines entirely arises in the hot core. 
In addition to the CH$_3$CN K-ladder we have also detected the CH$_3$$^{13}$CN
5$_0$$\rightarrow$4$_0$ line which allows us 
to have an estimate of the opacity of the CH$_3$CN line.

The CH$_3$CN column density has been estimated using the rotational
diagram technique. The low energy lines are expected to
be optically thick in the hot core. This is confirmed by our
interferometric observations. In particular,
we measure  I(CH$_3$CN 5$_0$$\rightarrow$4$_0$)/I(CH$_3$$^{13}$CN 5$_0$$\rightarrow$4$_0$)$\sim$2.4 
which implies an opacity $\sim$25 for  the main isotope line
assuming $^{12}$C/$^{13}$C=70. 
We have corrected by the opacity effect and derived
T$_{rot}$$\sim$54~K and a beam averaged column density
N(CH$_3$CN)$\sim$1.1~10$^{16}$~cm$^{-2}$
for the hot core component 
using the  5$_0$$\rightarrow$4$_0$  and  5$_5$$\rightarrow$4$_5$ lines.
The derived rotation temperature is in agreement, within the uncertainties,
 with that
derived from single dish data \citep{fue05}. Since 
the low energy lines are optically thick in the hot core, we think that this
agreement is fortuitous and due to the addition of two uncertainties 
which shift the estimated column
density in opposite directions, the contribution of extended emission in the
single-dish beam and the opacity effect.
From our interferometric observations, 
we obtain X(CH$_3$CN)$\sim$7.0~10$^{-9}$ 
in the  2~M$_\odot$ hot core in FIRS~2 assuming a source size of 0.72$"$$\times$0.52$"$ (see Table 5). 
This fractional abundance is similar to that measured in
massive hot cores and slightly larger than those derived in hot corinos.

For comparison, we have estimated the CH$_3$CN abundance
in the cooler and extended envelope. For this aim, we have   
subtracted the emission of the compact component from the single-dish spectrum and
analyze the result using a rotational diagram. The opacity effect
are expected to be less important in the low density envelope than in the hot
core. We derive
T$_{rot}$=12~K and a beam averaged column density of 5.8~10$^{11}$~cm$^{-2}$
(HPBW=25$''$). Using the H$_2$ column density derived from single-dish continuum
observations by Fuente et al. (2001), we estimate a
CH$_3$CN fractional abundance of $\sim$3~10$^{-11}$ in the extended
component, i.e. about 3 orders of magnitude lower than in the compact component.

Summarizing, our CH$_3$CN interferometric data show the existence of a compact 
(900$\times$650 AU) source characterized by a high kinetic gas temperature
($>$50~K) and enhanced CH$_3$CN abundance (X(CH$_3$CN)$\sim$7.0~10$^{-9}$)
in the Class 0 IM protostar FIRS~2. This constitutes a definite proof of the existence of a hot core
in this IM protostar which, up to our knowledge, it is the first IM hot core detected so far.  

\subsection{CH$_3$OH}
We have observed the CH$_3$OH J$_{k,k'}$=5$_{k,k'}$$\rightarrow$4$_{k,k'}$ lines towards
FIRS~2 using the 2.5 MHz wide band correlator which allows us
to observe most of the  CH$_3$OH J$_{k,k'}$=5$_{k,k'}$$\rightarrow$4$_{k,k'}$ components
simultaneously. In Fig. 1 we show the total integrated intensity map of the 
CH$_3$OH 5$_{1,4}$$\rightarrow$4$_{1,4}$ line which is that observed 
with the lowest upper stage energy (E$_u$=35 K). The emission of this component presents
two maxima, the first one is spatially coincident with the hot core while the second is located
at the position ($-2"$,$-1.5"$). This position lies on the axis of the outflow FIRS~2-out~1. 
In Fig 1 we also show
the integrated intensity map of the 5$_{3,1}$$\rightarrow$4$_{3,1}$,
5$_{3,2}$$\rightarrow$4$_{3,2}$,5$_{2,2}$$\rightarrow$4$_{2,2}$
and 5$_{3,3}$$\rightarrow$4$_{3,3}$ lines [panel indicated as CH$_3$OH (high)
in Fig.~1]. These lines are partially overlapped in our spectrum 
and all of them have upper state energies $>$50 K.
In this case, the emission is mainly arising in the hot core with a very weak contribution
of the second maximum. Thus, the morphology of the CH$_3$OH emission shows 
the existence of two components, a compact one which is spatially coincident
with the hot core (hereafter hot core component) 
and an extended one which is very likely associated with the outflow
FIRS~2--out~1 (hereafter outflow component). 

In Fig. 3 we compare the interferometric
spectra for the hot core and outflow components with
the single-dish spectrum at the central position.
The two components present differences in their kinematics and excitation
conditions.
The emission of the CH$_3$OH lines towards the hot core is centered at a
velocity $-$10$\pm$1~km~s$^{-1}$, while the emission 
of the outflow component is centered at $-$6$\pm$1~km~s$^{-1}$.
The $^{12}$CO~2$\rightarrow$1 and SiO~2$\rightarrow$1 maps of the
outflow FIRS~2--out~1 reported in Paper II show that
this velocity, -6$\pm$1~km~s$^{-1}$, is characteristic  
of the high velocity bullet R1,
reinforcing our interpretation of this emission as 
associated with the bipolar outflow. 

Different excitation conditions characterize the hot core and outflow components. 
This is clearly seen when one compares the relative intensities of 
the high and low energy lines in the methanol spectra towards the
two studied positions. 
All the CH$_3$OH lines have similar intensities towards the hot core.
However,  high energy lines are a factor of $\sim$5 weaker than the low energy 
ones towards the outflow position (see Fig. 3). We have
derived the rotation temperature and the beam averaged CH$_3$OH column density
for the hot core using the rotational diagram technique
(see Fig. 4). We obtain T$_{rot}$=809~K and 
N(CH$_3$OH)=1.5~10$^{17}$~cm$^{-2}$.
This temperature is quite high compared with that obtained from the
CH$_3$CN lines. 
Similarly to  the case of CH$_3$CN, the low energy lines are very likely
optically thick and the derived rotation temperature is an upper limit to
the actual one. To have an estimate of the uncertainty due to
the unknown rotation temperature in the CH$_3$OH column density, 
we have calculated the CH$_3$OH column density
from the flux of the 5$_{4,3}$$\rightarrow$4$_{4,3}$ line (E$_u$=104 K) 
assuming T$_{rot}$=50 K. We obtain N(CH$_3$OH)=4.9~10$^{16}$~cm$^{-2}$.
Thus, we consider that our {beam averaged column density estimate} 
is accurate within a factor of 3.
Assuming T$_{rot}$=50~K, the  derived CH$_3$OH fractional abundance 
is 3.0~10$^{-8}$ in the hot core. This abundance is
similar to that measured in the
prototypical massive hot core OMC1.

For comparison, we have estimated the rotation temperature and methanol
column density for the outflow condensation (see Fig.~4). We have obtained
T$_{rot}$=19~K and  N(CH$_3$OH)=2.9~10$^{15}$~cm$^{-2}$. The low rotational
temperature of this component supports our interpretation of being associated with
the outflow (postshocked material) instead of with the hot core. 
We have not detected a millimeter continuum
counterpart for this condensation in our interferometric 1mm map. In principle, 
this could be a sensitivity problem. From our interferometric image and
assuming T$_{dust}$=100~K we derive a 3$\times \sigma$ upper limit of $\sim$0.01~M$_\odot$
for the mass of the condensation associated with the methanol clump. This would imply
X(CH$_3$OH)$>$3.8~10$^{-7}$ in this clump. We can alternatively think that this clump is 
not the result of a hydrogen density enhancement at this position. 
For example,
it could be produced by a local enhancement of the CH$_3$OH abundance  
because of the shocks associated with the bipolar outflow.
To estimate the methanol abundance
in this case, we  calculate the H$_2$ column density from the 
single-dish $^{13}$CO observations
reported in Paper II. At the velocities of the bullet R1, the $^{13}$CO column  
density averaged in the 30m beam is $\sim$3~10$^{15}$~cm$^{-2}$. 
If we assume a uniform distribution of the
molecular gas in the beam, we obtain that the CH$_3$OH abundance
is similar to that of $^{13}$CO  for which we assume a standard  value of
$\sim$10$^{-6}$. Obviously, this is an upper limit to the actual abundance since we 
expect some clumpiness. Thus, the CH$_3$OH abundance in the outflow condensation
could be larger than in the hot core and reach values close to 10$^{-6}$.
This large value of the methanol abundance is consistent with those 
found in the molecular bullets associated
with low-mass bipolar outflows (see e.g. Bachiller \& P\'erez-Gutierrez 1997).

In addition to the CH$_3$OH lines, the spectrum
towards the hot core show several lines that are weaker towards
the outflow condensation. We have identified these lines as belonging to the complex
molecules CH$_3$OCH$_3$ and C$_2$H$_5$CN. We have also a tentative
detection of the sulphuretted species $^{34}$SO$_2$. The detection of
complex saturated molecules corroborates the existence of a  hot core  
in this source and reveals the rich and complex chemistry associated with it.

\setlength\unitlength{1cm}
\begin{figure}
\vspace{11cm}
\includegraphics{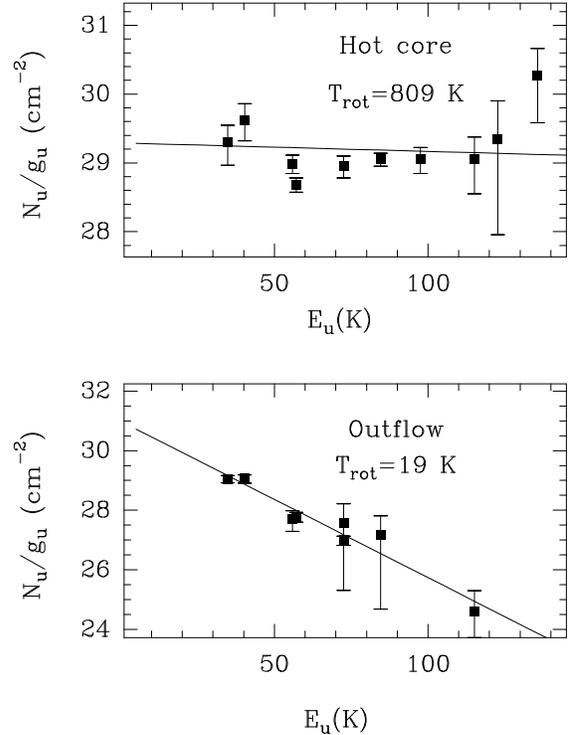}
\caption{Rotational diagram of CH$_3$OH towards the hot core and the outflow
condensation. Note that the rotation temperature towards the hot core is much larger
than towards the outflow condensation.}
\end{figure}

\subsection{D$_2$CO}
The interferometric map of D$_2$CO shows a compact source located at the position of
the hot core (see Fig.~1). The size of the D$_2$CO emitting region is similar to that
of CH$_3$CN. However,  the linewidth of the D$_2$CO line is $\sim$11~km~s$^{-1}$, 
larger than those of the CH$_3$CN and CH$_3$OH lines which are typically
$\sim$6--8~km~s$^{-1}$. Assuming the standard inside-outside collapse
model, the largest linewidth of the D$_2$CO line is consistent with the emission
arising in 
an inner region of the protostellar envelope. In a simple calculation,
an accretion velocity of $\sim$5~km~s$^{-1}$ corresponds to
a radius of $\sim$130~AU assuming a dynamical age of $\sim$10$^5$~yr and
dM/dt= 5~10$^{-5}$ M$_\odot$ yr$^{-1}$. A similar radius
was found by Maret et al. (2004) for the region where
the H$_2$CO abundance is heavily enhanced 
because of evaporation of the grain mantles in low-mass 
protostars. This suggests a similar origin for the D$_2$CO 
molecules in the hot core.

Our previous single dish spectrum was too noisy
to detect this wide emission (Paper II). 
Recently, we have taken a good signal-to-noise ratio spectrum to compare with
the interferometric observations (see Fig.~3).
The single-dish spectrum shows two components, a narrow one
at the velocity of the ambient cloud  (already detected in Paper II)
which is completely missed in our interferometric observations
and  a wide one with a linewidth similar to that of 
the PdB spectrum . Even in this wide component the interferometer
recovers only the 50\% of the flux, suggesting that
part of them arises in a more extended area.

A narrow and wide components with linewidths similar to those of
D$_2$CO are also detected in the single dish spectra of the 
H$_2$CO lines (Paper II). In Paper II, we have separately mapped
the narrow and wide components.
These maps revealed that both components are 
associated with the outflow. While the wide component arises in the
jet, the emission of the narrow component was more intense in the
interface between the jet and the molecular cloud suggesting that it is 
tracing the molecular gas that is being swept up by the 
jet. We propose that
the missed flux in our interferometric D$_2$CO image is probably
arising in a more extended component associated with the bipolar 
outflow similarly to the case of H$_2$CO. 

Our interferometric image 
shows that, in addition to the extended component, there is 
an intense and compact D$_2$CO component associated
with the hot core.
We have estimated the D$_2$CO abundance in this component
assuming T$_{rot}$=50~K and obtained 
N(D$_2$CO)=3.5~10$^{14}$~cm$^{-2}$ and X(D$_2$CO)$\sim$1.4~10$^{-10}$. 
This fractional abundance is a factor of $\sim$2 higher than that derived 
in Paper II for the whole envelope.

\subsection{N$_2$D$^+$} 
We have not detected the N$_2$D$^+$ 3$\rightarrow$2 line 
in our interferometric image. 
The rms of the image is 50 mJy/beam which is about $\sim$2\% of the 
flux measured with the 30m telescope (Paper II). Thus, almost the 100\% of the
N$_2$D$^+$ emission has been missed in the interferometric observations. 
This implies that the emission of N$_2$D$^+$ presents a quite uniform
distribution across the envelope and suggests that it mainly arises in the 
cooler extended 
envelope. This is also in agreement with our interpretation in Paper II
based on kinematical arguments. The linewidth of the N$_2$D$^+$ 
3$\rightarrow$2 line, $\sim$1~km~s$^{-1}$, is a at least a
factor of 2--3 smaller than those of the lines arising in the warm component,
and similar to those of the molecules which are thought to be good tracers of the cold
envelope such as H$^{13}$CO$^+$ and N$_2$H$^+$.  

\setlength\unitlength{1cm}
\begin{figure*}
\vspace{20cm}
\includegraphics{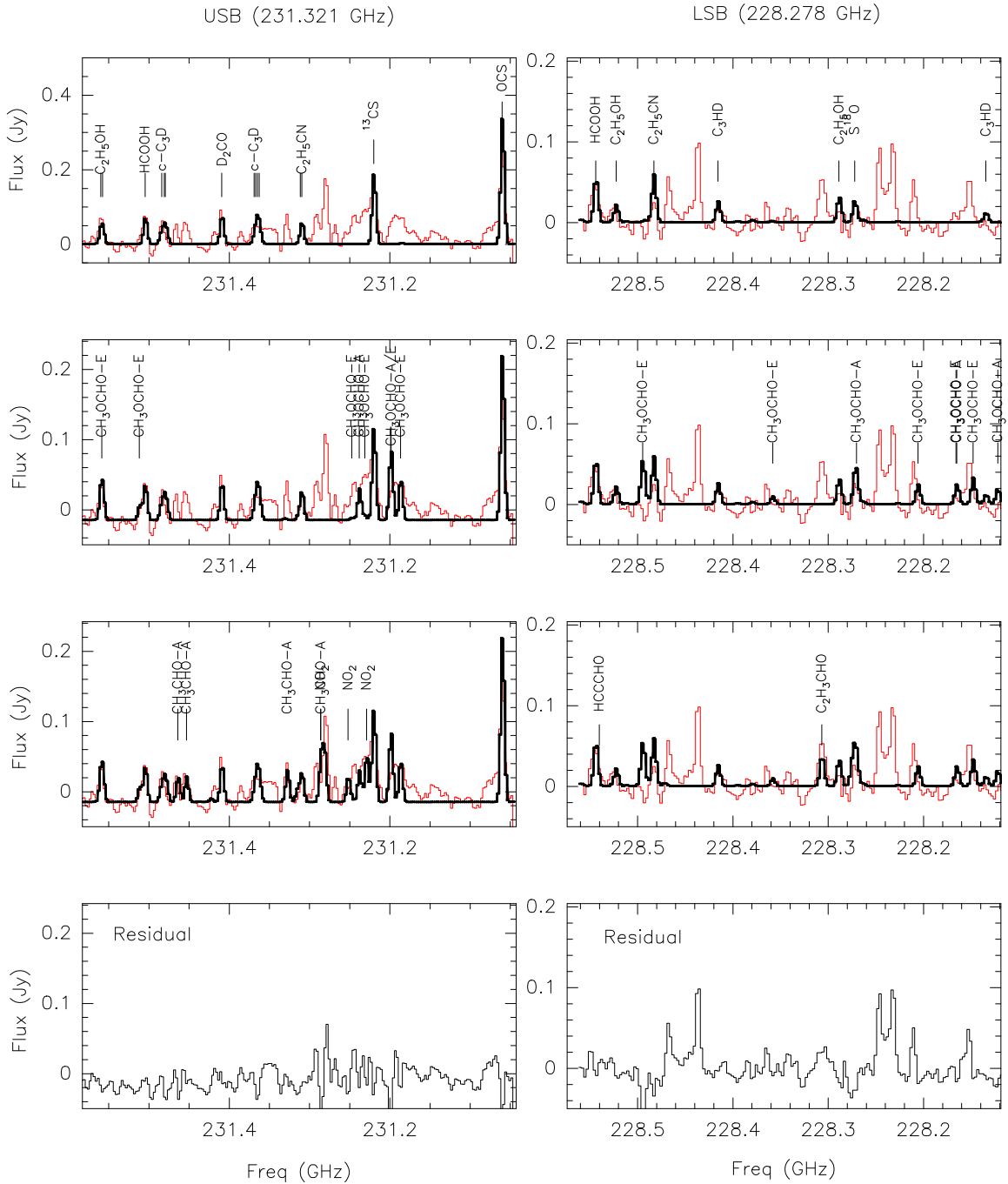}
\caption{
Spectra of the upper side band (USB) and lower side band (LSB) of the 1mm receiver obtained in our
PdBI observations with our best guess for line identifications. 
For clarity, we have divided the molecular species in three groups. 
The first group is formed by the more reliable identifications and S$^{18}$O.
Beginning from the top, in the first line of panels we compare the observed spectra
with the synthesized ones taking into account only this first group.  
In the second line, we add CH$_3$OCHO-A /E
and show the obtained synthesized spectra. The inclusion of these compounds improve the
agreement with the observational data. However, the large CH$_3$OCHO-A /E abundance
we derive from these observations, rise some questions about this identification. 
In the third line of panels, we show the synthesized
spectra after adding some other exotic compounds. In the last line, it is shown the
residual spectra after subtracting our fit. Note that the agreement between the
synthesized and observed one is very good in the USB but some
lines remain unidentified in the LSB.}
\end{figure*}

\begin{table}
{\scriptsize
\begin{center}
\caption{USB - 2.4 MHz \label{tbl-2}}
\begin{tabular}{llccc}
\hline\hline
\multicolumn{1}{c}{Freq.} &
\multicolumn{1}{c}{Molecule} & 
\multicolumn{1}{c}{Transition} &
\multicolumn{1}{c}{Int$^a$} & 
\multicolumn{1}{c}{E$_u$(K)}  \\
\hline
{\bf 231060.98} & {\bf OCS}   & {\bf $19 \rightarrow 18$} & {\bf -2.6263} & {\bf 100} \\
231187.69 &  CH$_3$OCHO-E  & $21_{9,13} \rightarrow 21_{8,14}$ &   -4.8522 & 179 \\
231199.96 & CH$_3$OCHO-A  & $21_{9,12} \rightarrow 21_{8,13}$  & -4.8484 & 179 \\
231200.14 &  CH$_3$OCHO-E   & $21_{9,12} \rightarrow 21_{8,13}$  & -4.8522 & 179 \\
{\bf 231220.99} & {\bf $^{13}$CS}  &  {\bf $5 \rightarrow 4$}   &{\bf  -4.2103} & {\bf 51} \\
231229.96 & NO$_2$ & $4_{1,3} \rightarrow 4_{0,4}$ &  -4.8527  &    12 \\   
231239.07 &  CH$_3$OCHO-A & $21_{9,13} \rightarrow 21_{8,14}$ & -4.8482 & 179 \\
231232.06 &  CH$_3$OCHO-E & $29_{6,24} \rightarrow 29_{5,25}$ & -5.0160 & 253 \\
231248.31 &  CH$_3$OCHO-E & $29_{6,24} \rightarrow 29_{4,26}$  & -5.3013 & 253 \\
231252.98  & NO$_2$                & $4_{1,3} \rightarrow 4_{0,4}$  & -5.0896 &   12  \\   
231279.18  & CH$_3$CHO-A    & $12_{6,7} \rightarrow 11_{6,6}$  &  -2.9783  &   142 \\
231279.18   & CH$_3$CHO-A    & $12_{6,6} \rightarrow 11_{6,5}$  & -2.9783  &   142 \\
231286.40          & CH$_3$OH     & $10_{2,2} \rightarrow 9_{3,2}$   & -4.0689           &  154  \\
231287.14 & NO$_2$ &   $4_{1,3} \rightarrow 4_{0,4}$  & -5.0875  &   12 \\   
{\bf 231310.42} &  {\bf C$_2$H$_5$CN} & {\bf $26_{1,25} \rightarrow 25_{1,24}$}  & {\bf -2.6517} &   {\bf 142} \\
231320.82        &  CH$_3$CHO-A    &  $12_{5,8} \rightarrow 11_{5,7}$    &   -2.9001    &  117 \\
231320.97         & CH$_3$CHO-A    &  $12_{5,7} \rightarrow  11_{5,6}$    &   -2.9001     & 117  \\
{\bf 231365.80}  &  {\bf c-C$_3$D} & {\bf $5_{3,3} \rightarrow 4_{3,2}$}      &{\bf -3.1276} & {\bf 28} \\
{\bf 231410.27}  & {\bf D$_2$CO} & $4_{0,4} \rightarrow 4_{0,3}$   &   {\bf -2.4242} & {\bf 17} \\
231446.49  &  CH$_3$CHO-A & $12_{4,9} \rightarrow 11_{4,8}$  &  -2.8386 & 97 \\
231457.25  &  CH$_3$CHO-A & $12_{4,8} \rightarrow 11_{4,7}$  &  -2.8386 & 97 \\
{\bf 231485.11} &  {\bf c-C$_3$D} & $5_{3,3} \rightarrow 4_{3,2}$  &  {\bf -3.4082} & {\bf 28} \\
{\bf 231505.70} &  {\bf HCOOH} & $10_{1,9} \rightarrow 9_{1,8}$   &{\bf -3.1963} & {\bf 53} \\
231511.81  &  CH$_3$OCHO-E &  $27_{4,24} \rightarrow 27_{2,25}$  &  -5.3849  &  211 \\
231513.06  &  CH$_3$OCHO-E  &  $27_{4,24} \rightarrow 27_{3,25}$  & -5.1294  &  211 \\
{\bf 231558.55}  & {\bf C$_2$H$_5$OH}  &  {\bf $21_{5,17} \rightarrow 21_{4,18}$}  &  {\bf -3.6252}  &  {\bf 215} \\
231559.62  &  CH$_3$OCHO-E  &  $27_{5,23} \rightarrow 27_{2,25}$  & -5.1292  &  211 \\
{\bf 231560.90} &  {\bf C$_2$H$_5$OH} & {\bf $20_{5,16} \rightarrow 20_{4,17}$}  & {\bf -3.6250} & {\bf 197}  \\
\hline
\end{tabular}
\end{center}
$^a$ Base 10 logarithm of the integrated intensity in units of nm$^2$-MHz as appears tabulated in the JPL
molecular line catalogue (Pickett et al. 1998).
}
\end{table}

\begin{table}
{\scriptsize
\begin{center}
\caption{LSB - 2.4 MHz \label{tbl-2}}
\begin{tabular}{llccc}
\hline\hline
\multicolumn{1}{c}{Freq.} &
\multicolumn{1}{c}{Molecule} & \multicolumn{1}{c}{Transition} &
\multicolumn{1}{c}{Int$^a$} & 
\multicolumn{1}{c}{E$_u$(K)}  \\
\hline
 228122.07       &  CH$_3$OCHO-A         & $5_{4,2} \rightarrow 4_{2,3}$  &  -7.1993       &  8 \\
{\bf 228134.42} & {\bf c-C$_3$HD}           &  {\bf $15_{7,8} \rightarrow 15_{7,9}$}    &   {\bf -3.9965} &  {\bf 288} \\
228147.97         &  CH$_3$OCHO-E        &  $18_{6,13} \rightarrow 18_{2,16}$        &    -5.9576        &   114 \\
228165.01         &  CH$_3$OCHO-A        &  $18_{6,13} \rightarrow 18_{4,14}$         &    -5.9537         &  114 \\
228165.86         &  CH$_3$OCHO-E        &  $44_{14,31} \rightarrow 43_{11,32}$     &    -5.9853         &  673 \\
228270.42         &  CH$_3$OCHO-A         &  $24_{9,16} \rightarrow 24_{8,17}$        &    -4.8334         &   220 \\
228272.26         &  S$^ {18}$O                   &                $9_8 \rightarrow 8_8$             &     -3.9573          &  82 \\ 
{\bf 228288.69}  & {\bf C$_2$H$_5$OH}   &  {\bf $11_{11,0} \rightarrow 10_{2,8}$}  &   {\bf -4.0647}   &  {\bf 107} \\
228205.84          & CH$_3$OCHO-E        &  $24_{9,16} \rightarrow 24_{8,17}$              &        -4.9577           &  220 \\
228232               & Unidentified                  &       &                   &         \\
228245               & Unidentified                  &        &                  &         \\ 
228278.96          & HCCCHO                    &  $25_{1,25} \rightarrow 24_{1,24}$  &    -2.8001           &  135 \\
228310.56          & C$_2$H$_3$CHO       &  $13_{2,12} \rightarrow 12_{1,11}$   &   -4.6934           &  36 \\
228358.20          & CH$_3$OCHO-E        & $24_{9,15} \rightarrow 24_{8,17}$   &   -5.4380         &  220 \\
 {\bf 228415.69} & {\bf c-C$_3$HD} &  $11_{8,4} \rightarrow 11_{6,5}$ &  {\bf -4.5197}  &  {\bf 169} \\
228434               & Unidentified                 &       &                &         \\
228467               & Unidentified                 &        &               &          \\
 {\bf 228483.13} & {\bf C$_2$H$_5$CN} &   $25_{2,23} \rightarrow 24_{2,22}$  & {\bf -5.4674} &  {\bf 158} \\
  228494.88       & CH$_3$OCHO-E       &  $14_{4,10} \rightarrow 14_{1,13}$    &   -6.2908 3      & 62 \\
 {\bf 228522.71}         & {\bf C$_2$H$_5$OH}  &  {\bf $7_{3,4} \rightarrow 7_{2,6}$}  &  {\bf -4.3308}  &  {\bf 84} \\
 {\bf 228544.17}         & {\bf HCOOH} & {\bf $10_{2,8} \rightarrow 9_{2,7}$} & {\bf -3.2324} &  {\bf 62} \\
\hline
\end{tabular}
\end{center}
$^a$ The same as in Table 3.
}
\end{table}

\begin{table*}
\begin{center}
\caption{Fractional Abundances in the hot core of NGC 7129--FIRS 2 and comparison
with other hot cores \label{tbl-3}}
\begin{tabular}{lccccccc}
\hline \hline
\multicolumn{1}{c}{} &
\multicolumn{3}{c}{NGC 7129--FIRS 2} & 
\multicolumn{1}{c}{NGC 1333 IRAS4 A} &
\multicolumn{1}{c}{IRAS 16293} &
\multicolumn{1}{c}{OMC 1}  &  \multicolumn{1}{c}{G327.3-0.6}\\
 & \multicolumn{3}{c}{$\sim$ 500 L$_{\odot}$} & 
\multicolumn{1}{c}{$\sim$ 14 L$_{\odot}$} & 
\multicolumn{1}{c}{$\sim$ 23 L$_{\odot}$}  &  
\multicolumn{1}{c}{$\sim$1.0 10$^4$ L$_{\odot}$}  & 
\multicolumn{1}{c}{$\sim$ 10$^5$ L$_{\odot}$} \\
\multicolumn{1}{c}{Molecule} &
\multicolumn{1}{c}{T$_{rot}$ (K)} & \multicolumn{1}{c}{N$^*$(cm$^{-2}$)}   &
\multicolumn{1}{c}{X$^a$}  &  \multicolumn{1}{c}{X$^b$} &
\multicolumn{1}{c}{X$^c$}   & \multicolumn{1}{c}{X$^d$}  &  \multicolumn{1}{c}{X$^e$} \\
\hline
H$_2$             &                    & 7.9$\times$10$^{24}$  &    1   &  1   &  1  &   1   & 1\\
CH$_3$CN     &         54        &  5.5$\times$10$^{16}$   &   7.0$\times$10$^{-9}$   &   
1.6$\times$10$^{-9}$  & 1.0$\times$10$^{-8}$ &  4$\times$10$^{-9}$  & 7$\times$10$^{-7}$  \\
CH$_3$OH     &       50$^a$  &  2.4$\times$10$^{17}$   &   3.0$\times$10$^{-8}$   &   
  $<$7 10$^{-9}$ & 3.0$\times$10$^{-7}$$^e$  & 1$\times$10$^{-7}$  & 2$\times$10$^{-5}$ \\
H$_2$CO       &  50$^a$  &  2.5$\times$10$^{15}$   &   3.1$\times$10$^{-10}$  &
2.0$\times$10$^{-8}$  & 6.0$\times$10$^{-8}$   &  7.0$\times$10$^{-9}$  &      \\
D$_2$CO        &       50$^a$  &  3.5$\times$10$^{14}$   &   4.4$\times$10$^{-11}$  &  
     &   3.0$\times$10$^{-9}$$^{g}$    &          &                   \\
$^{13}$CS       &       50$^a$   & 3.4$\times$10$^{14}$   &   4.3$\times$10$^{-11}$     & 
                                      &   & 7$\times$10$^{-11}$  &    5$\times$10$^{-11}$$^g$      \\
OCS                 &      50$^a$    & 2.4$\times$10$^{16}$   &  3.0$\times$10$^{-9}$    &  
                        & 1.0$\times$10$^{-6}$   & 7$\times$10$^{-8}$   &   2$\times$10$^{-8}$$^g$       \\
HCOOH           &      48            & 3.5$\times$10$^{15}$   &  4.4$\times$10$^{-10}$   &    
 4.6$\times$10$^{-9}$  &  6.2$\times$10$^{-8}$  & 8$\times$10$^{-10}$    & 3$\times$10$^{-10}$$^h$            \\
C$_2$H$_5$OH    & 75   &  2.0$\times$10$^{16}$  &  2.5$\times$10$^{-9}$  &            
    &    &  7$\times$10$^{-10}$ & 3$\times$10$^{-9}$$^h$  \\
C$_2$H$_5$CN   & 50$^a$   &  3.2$\times$10$^{15}$  & 4.0$\times$10$^{-10}$ &
 $<$1.2$\times$10$^{-9}$  & 1.2$\times$10$^{-8}$   & 3.0$\times$10$^{-9}$ &  4$\times$10$^{-7}$ \\
c-C$_3$D         &     50$^a$     & 2.5$\times$10$^{15}$   &  3.1$\times$10$^{-10}$   &        &               &               &                                   \\
c-C$_3$HD      &      50           &  1.8$\times$10$^{17}$    &  2.3$\times$10$^{-8}$   &         &              &              &                                   \\
 \hline
\multicolumn{8}{c}{Uncertain identifications} \\
S$^{18}$O       &       50$^a$   & 2.6$\times$10$^{16}$   &   3.3$\times$10$^{-9}$    &  
   & 2.6$\times$10$^{-9}$      &   3$\times$10$^{-10}$    &   $>$6$\times$10$^{-13}$      \\
CH$_3$OCHO-E    & 36  & 4.0$\times$10$^{18}$  &  5.0$\times$10$^{-7}$   & 
 3.6$\times$10$^{-8}$ &  2.3$\times$10$^{-7}$  &1$\times$10$^{-8}$  &   2$\times$10$^{-6}$   \\
CH$_3$OCHO-A  & 36$^a$    & 3.3$\times$10$^{18}$   & 4.1$\times$10$^{-7}$   & 
 3.4$\times$10$^{-8}$  &1.7$\times$10$^{-7}$ & 1$\times$10$^{-8}$  & 2$\times$10$^{-6}$ \\ 
CH$_3$CHO-A    &   50$^a$  & 2.3$\times$10$^{15}$  &  2.9$\times$10$^{-10}$  & 
   & 1.9$\times$10$^{-8}$   &    & 5$\times$10$^{-10}$$^g$  \\
HCCCHO    & 50$^a$  & 2.0$\times$10$^{15}$   &  2.5$\times$10$^{-10}$  &
                    &                &   &     \\
C$_2$H$_3$CHO  &  50$^a$ & 5.0$\times$10$^{16}$  & 6.3$\times$10$^{-9}$  &
                     &              &     &      \\
CH$_3$OCH$_3$ & 50$^a$ & 1.9$\times$10$^{15}$ &  2.4$\times$10$^{-10}$  &
 $<$2.8$\times$10$^{-8}$   &  2.4$\times$10$^{-7}$    & 8$\times$10$^{-9}$  & 3$\times$10$^{-8}$$^g$    \\
$^{34}$SO$_2$  & 50$^a$ & 3.8$\times$10$^{13}$ &  4.8$\times$10$^{-12}$  &
              &   2.4$\times$10$^{-8}$     & 2$\times$10$^{-10}$    & 4$\times$10$^{-9}$$^g$     \\
NO$_2$       & 50$^a$   & 1.0$\times$10$^{17}$  &  1.2$\times$10$^{-8}$ &
                    &             &      &        \\ \hline
\end{tabular}
\end{center}
$^*$ Values derived for the NGC 7129--FIRS 2 hot core assuming a size of $\sim$0.72$"$$\times$0.52$"$. 
To derive the H$_2$ column density we have considered that the hot core mass (2 M$_\odot$) is uniformly 
distributed in an area $\sim$$a b$ with $a$=0.004 pc (0.72$"$)  and $b$=0.003 pc (0.52$"$).\\
$^a$ Assumed temperature.\\
$^b$ Values from Bottinelli et al. (2004) for a 0.0016 pc (0.5$''$) source.\\
$^c$ Values from Cazaux et al. (2003) and Wakelam et al. (2004) for a 0.0016 pc (2$''$) source. \\
$^d$ Values from Sutton et al. (1995) for a 0.034 pc (14$''$) source.\\
$^e$ Values from Gibb et al. (2000) for a 0.03 pc (2$''$) source.\\
$^f$ From Schoier et al. (2002).\\
$^g$ From  Ceccarelli et al. (2001).\\
$^h$ From Gibb et al. (2000) for a 0.32 pc (20$''$) source.\\
\end{table*}

\section{Chemical complexity of the hot core in NGC~7129--FIRS~2}
In Fig.~5 we show the observed wide band spectra in the 
upper side band (USB) and lower side band (LSB) of the 
1mm receiver. We have detected a real forest of lines in the 1mm band. 
The identification of these lines is not an 
easy task. Since we have moderate spectral resolution (2.5 MHz), several
candidates can be found for each line. On the other hand, in most cases we have
two or more lines overlapped. We have tried to identify most of them with
the following procedure: In a first step we select our best candidates 
for the lines and  estimate the column density of the carriers
assuming optically thin emission and local thermodynamic equilibrium (LTE).
In a few cases, we are able to estimate the rotation temperature because we 
have at least two $``$isolated" lines (see Table 3). In the other cases, 
we have assumed T$_{rot}$=50~K. In a second step, we  synthesized the expected spectrum 
assuming  the estimated column densities. The velocity profile is assumed to be Gaussian
with a half power full width of 7~km~s$^{-1}$ and centered at the ambient cloud
velocity, -10 km s$^{-1}$. Finally, we compare it with the observed 
spectrum and readjust the molecular column densities to improve the fit.
We list in Tables~3 and 4 the lines detected in the USB and LSB spectra with
the most likely line identification. 

Obviously, this identification procedure entails some ambiguity. We have classified
the species according with the reliability of their identification in two groups that
are differentiated in Table~5. The first group contains the most reliable 
identifications. In these cases, the observed frequencies  
are in total agreement with those calculated for these species. 
Moreover, the synthesized spectrum
matches quite well with the observed one (see top panel of Fig.~5). 
Finally, the derived fractional abundances
are within the range of abundances measured for these
species in hot cores. 
Within this group, we have OCS, $^{13}$CS, C$_2$H$_5$CN, c-C$_3$D, D$_2$CO, HCOOH, and
C$_2$H$_5$OH in the USB and C$_2$H$_5$OH, C$_3$HD in the 
LSB. The lines of the first group species are indicated by bold characters 
 in Tables~3 and 4. The derived column densities
and fractional abundances are shown in Table~5.

We have put the complex molecules CH$_3$OCHO-A and CH$_3$OCHO-E (methyl formate) 
in the second group. Several lines of these species have been detected in the USB 
and LSB (see second panels in Fig.~5).  
However, the derived fractional abundances in NGC 7129-FIRS 2 are unexpectedly high
compared to those measured in hot corinos and massive hot cores (see Table~3). 
One possibility is that the
rotation temperature is higher than that assumed. To estimate the uncertainty
due to the poorly known rotation temperature, we have repeated the column
density calculations with T$_{rot}$=300~K. The derived abundance is an order
of magnitude lower and then it agrees with
that derived in IRAS 16293. This high rotation temperature could be
explained if this complex
molecule arises in an inner and hotter region than CH$_3$CN. 
If methyl formate cannot be produced in the gas phase (as 
suggested by the laboratory results of Horn et al. 2004), our findings
may imply that this species is formed on grain surfaces and that its 
binding energy is larger than that of CH$_3$CN, thus requiring  
larger dust temperatures to desorb.

The rarer isotope S$^{18}$O has also been included in the group of uncertain 
identifications. We have only one line of this molecule,
S$^{18}$O 9$_8$$\rightarrow$8$_8$ at 228.272 GHz, and the derived S$^{18}$O abundance
is unexpectedly high. In this case we could have a misidentification. 
There is a CH$_3$OCHO-A line
at 228.270 GHz, that could be an alternative
identification of  the tentative S$^{18}$O line if the methyl formate  abundance 
is unexpectedly high. 

The most uncertain identifications are C$_2$H$_3$CHO, CH$_3$CHO-A, HCCCHO, 
and NO$_2$. The observed frequencies are in agreement with those of C$_2$H$_3$CHO,
CH$_3$CHO-A,  and HCCCHO lines if the emission arises in gas at a velocity
v$_{lsr}$$\approx$$-$5~km~s$^{-1}$. This velocity is coincident with
that of the  bullet R1 (see Paper I) but is shifted by $\sim$5~km~s$^{-1}$
from that of the molecular cloud. Similarly, the fit to the USB spectrum improves if we include
a high excitation methanol line (see Table~3) 
emitting at the velocities $-$5 and $-$20~km~s$^{-1}$. These velocities are within
the range of velocities associated with this Class 0 protostar but
are shifted from that of the ambient cloud. 
NO$_2$ is not a common molecule in hot cores, 
and the three lines observed in the spectrum are overlapped with others of more common
hot core species. We have kept it in the table  because
its inclusion improves the fit to the observed spectrum.

In Fig.~5  we show the synthesized USB and LSB spectra
assuming the molecular column densities listed in Table~3. The 
fit is quite good for the USB. However, it is poor for the LSB where
we still have 4 unidentified lines. This is quite frustrating taking into account
that they are the most intense lines in this subband. The frequencies of
the unidentified lines are 228467($\pm$5), 228434($\pm$5),228245($\pm$5)
and 228232($\pm$5) MHz, and we do not have any good candidates
for them. There is a line of vibrationally excited DC$_3$N at the frequency
of 228467.44 GHz. Emission of vibrationally excited HC$_3$N has been 
detected in massive hot cores (see e.g. Wyrowski et al. 1999). 
However, if this identification were correct, we would expect to see
a vibrationally excited HC$_3$N line at least ten times more intense at
a frequency of 228303 MHz which is not detected. For the line at 228434 MHz our best
candidate is glycoaldehyde (CH$_2$(OH)CHO). Glycoaldehyde has already 
been observed in the interstellar medium \citep{hol04}. However, the upper state energy of
the line at 228434 MHz is too high ($>$1400 K) to be detectable in this source.
The line at 228245 MHz could be propadienone (CH$_2$CCO), but in this case we should
have detected other lines of this compound in the same spectrum. Finally, our best 
candidate for the line at  228232 MHz is cyclopropenone (C$_3$H$_2$O). 
This molecule has never been detected in space. Since we have only
one observable line in our spectra, we cannot confirm this detection.

In addition to the lines detected in the wide-band spectrum shown in 
Fig.~5, in Table~5 we also include the column densities derived for
CH$_3$OCH$_3$ and $^{34}$SO$_2$, these species detected 
in the CH$_3$OH spectrum 
shown in Fig.~3.  Since we have only one
line for these species, the derived column density is also uncertain. 

\section{Discussion}

In Table~5 we list the derived abundances of the molecules detected in 
the hot core towards FIRS 2. A first look at the Table  reveals that
FIRS~2 is rich in deuterated species (D$_2$CO,c-C$_3$D
and c-C$_3$HD), sulphuretted compounds ($^{13}$CS, OCS) 
and complex O- and N- molecules (HCOOH,
C$_2$H$_5$OH, C$_2$H$_5$CN). Taking into account the large
uncertainty in the estimated fractional abundances, we have centered
our discussion in the most reliable identifications, i.e., the first group 
in Table 3.

\subsection{Deuterated species}
The deuterium fractionation of molecular species is a well 
known process in dense interstellar clouds (Watson, W.D. 1974). 
In the gas phase, this is mainly driven by the enhancement 
of the H$_2$D$^+$/H$_3^+$ ratio due to the exothermic 
proton--deuteron exchange reaction H$_3^+$ + HD $\rightarrow$
H$_2$D$^+$ + H$_2$ (+ 230 K), which does not proceed backwards if the 
gas temperature is less than about 30 K.  This process is 
responsible for the deuteration of species such as HCO$^+$ and 
N$_2$H$^+$, via the reaction of H$_2$D$^+$ with the parent species
CO and N$_2$, respectively, as observed in dense cloud cores 
\citep{but95,wil98}.  

Another phenomenon which boosts the deuterium 
fractionation is the freeze out of neutral species, in particular the abundant
O and CO, which are efficient destruction partners of  
H$_3^+$ and H$_2$D$^+$ in the gas phase. Their freeze out, known to 
happen in dense cold clouds  \citep{wil98,cas99,kra99,ber02,taf02},
further enhances the H$_2$D$^+$/H$_3^+$ abundance ratio (because
of the enhanced formation rate of H$_2$D$^+$, and the reduced 
destruction rates for H$_2$D$^+$ and H$_3^+$) and the consequent 
transfer of the deuteron to gas-phase species \citep{cas02,bac03,cra05}. 
Once CO is highly 
depleted, multiple deuterated forms of H$_3^+$ can easily be produced 
\citep{rob03,wal04}. Indeed, large abundances
of H$_2$D$^+$ and D$_2$H$^+$ have been measured toward two 
pre--stellar cores \citep{cas03,vas04}.
This has the consequence of enhancing the gaseous D/H ratio 
to values around 0.1, four orders of magnitude larger than the cosmic D 
abundance, which allows an efficient deuteration on grain 
surfaces \citep{tie83,cha97,sta03}. 

We have detected the deuterated species D$_2$CO, c-C$_3$D and
c-C$_3$HD towards FIRS~2. Moreover, the D$_2$CO
interferometric image shows that the D$_2$CO emission is strongly
peaked towards the hot core position. Recent theoretical and
observational research suggests that 
the abundance of the non-deuterated species H$_2$CO is enhanced 
by more than 2 orders of magnitude in the inner 
region of the envelope, where the dust temperature is larger than 100~K
\citep{mar04}. The size of this region ($\sim$100~AU) is similar
to that derived for the D$_2$CO emission from the linewidth of the D$_2$CO line.
Assuming that evaporation is also the origin of D$_2$CO, 
the [D$_2$CO]/[H$_2$CO] would tell us about the gaseous D/H ratio
during the pre-stellar phase of the parent clump. 
Unfortunately, we have not interferometric observations of
H$_2$CO and cannot derive an accurate [D$_2$CO]/[H$_2$CO]
ratio in the hot core. Trying to get an estimate of the deuteration
degree of formaldehyde, we have compared the single dish spectra
of the D$_2$CO 4$_{04}$$\rightarrow$3$_{03}$ 
and H$_2$CO 3$_{12}$$\rightarrow$2$_{11}$ lines (see Fig. 6). Both lines have
similar excitation conditions and frequency. 
The line intensity ratio in the narrow component is
lower by a factor of 2 than in the wide component suggesting a
higher [D$_2$CO]/[H$_2$CO] ratio in the hot core. Assuming
the same physical conditions for D$_2$CO and H$_2$CO we
derive [D$_2$CO]/[H$_2$CO]$\sim$0.14$\pm$0.03 in the hot core.
This value is similar to that found in pre-stellar cores
and low-mass protostars \citep{cas02,bac03,cec01,loi03}
and it implies large D/H ratios during
the cold pre-stellar phase (D/H$\sim$0.5, see Caselli et al. 2002b;
Stantcheva \& Herbst 2003).

Contrary to D$_2$CO, we have not detected
N$_2$D$^+$ in our interferometric observations. 
It is interesting to note that whereas the deuteration of N$_2$H$^+$ 
is a gas--phase process (which proceeds from the reaction between 
N$_2$ and H$_2$D$^+$),  the formation of D$_2$CO seems 
to require surface chemistry \citep{cec02}. Therefore,
our result is not surprising.  In fact, as stated above, 
the deuteration in the gas phase is efficient only if the gas temperature 
is lower than, say, 30 K.  Thus, in a hot core, we do not expect to 
see N$_2$D$^+$.  Moreover, in hot cores, where icy grain mantles 
evaporates, CO and H$_2$O molecules are abundant in the gas
phase and they efficiently destroy 
N$_2$H$^+$ (and N$_2$D$^+$), so that one does not expect to see 
large abundances of N$_2$H$^+$ either. 
The non-detection of N$_2$D$^+$ in the hot core associated with
FIRS~2 confirms the interpretation given in Paper II
that N$_2$D$^+$ mainly arises in the cool envelope,
and constitutes an indirect proof of this chemical scheme.  On the other hand, the large 
D$_2$CO/H$_2$CO abundance ratio observed in the same position
can be easily explained if the deuterium fractionation of formaldehyde 
happened on grain surfaces during the cold pre--stellar phase, and if 
icy mantles have recently evaporated, so 
that gas phase chemistry did not have the time to significantly
reduce the abundance of the deuterated ``parent'' species.

\setlength\unitlength{1cm}
\begin{figure}
\vspace{12cm}
\includegraphics{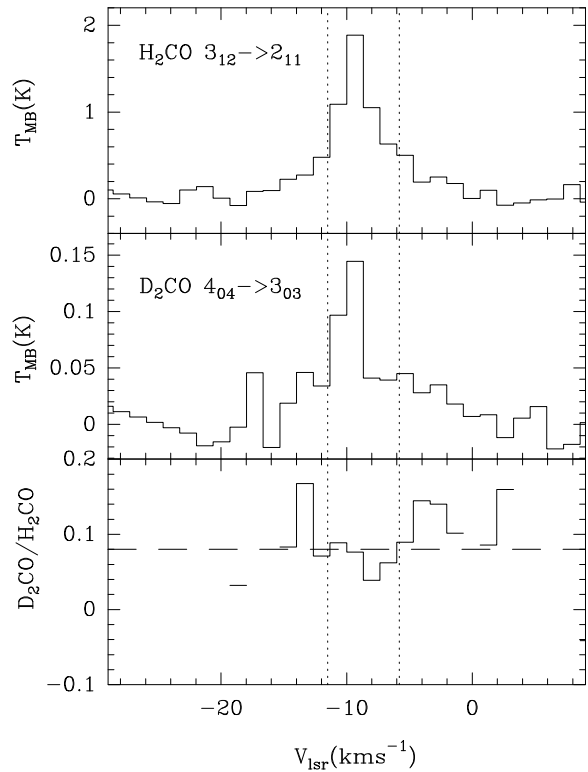}
\caption{Single-dish spectra of the H$_2$CO 3$_{12}$$\rightarrow$2$_{11}$ and
D$_2$CO 4$_{04}$$\rightarrow$3$_{03}$ lines towards FIRS~2. In the bottom panel
we show the I(D$_2$CO 4$_{04}$$\rightarrow$3$_{03}$)/I(H$_2$CO 3$_{12}$$\rightarrow$2$_{11}$)
line ratio. Note that this ratio presents a minimum around the velocity of the ambient cloud
($\sim$$-$10~km~s$^{-1}$). }
\end{figure}

\subsection{Sulphuretted species}
The abundances of sulphuretted molecules are very dependent on the
stellar age. We have compared the observed [CS]/[OCS] ratio in FIRS 2 
($\sim$1.1 assuming CS/$^{13}$CS=89) with the Viti et al. (2004) model for 
a 5~M$_\odot$ star. The measured ratio is consistent with 
the FIRS 2 being an evolved hot core with an age of $>$1.0~10$^5$ yr.
The detection of large species like C$_2$H$_5$OH
also suggests that we are dealing with an evolved hot core since these
species are only formed at late times. Nevertheless, we would like to
point out that the determination of the hot core age using the 
sulphuretted molecules is quite uncertain since other 
factors like the ice mantle composition and the hot core kinetic temperature 
can also strongly affect the abundances of these species (Wakelam et al. 2004).

\subsection{Complex O- and N-bearing molecules}

Theoretical models predict that the relative abundance of 
the N-bearing and O-bearing complex molecules is dependent on the hot 
core temperature  \citep{cas93,rod01}.
The luminosity of FIRS~2 is a factor $>$10 larger than those of the two 
low mass hot corinos detected so far and a factor
$\sim$200 lower than that of the massive protostar creating the Orion hot core. 
The intermediate kinetic temperature of the FIRS~2 hot core is expected to 
produce a differentiated chemistry of complex molecules. 
In order to have a deeper insight into the dependence of the complex molecules
abundances on the protostellar luminosity, in Table~5
we compare the observed abundances towards FIRS~2
with those measured in prototypical hot corinos and massive hot cores.
The luminosity of the objects listed in Table~5 varies by 4 orders of 
magnitude, thus one expects to detect some chemical differences
between these hot cores.

However, the comparison of abundances between different hot cores (corinos)
is difficult. First of all, since massive stars are usually located farther
from the Sun than their low- and intermediate-mass counterparts, 
the abundances are derived in different spatial
scales. The fractional abundances shown in Table 5
are averaged values in regions with sizes ranging 
from $\sim$0.002~pc in NGC1333--IRAS~4A to  
0.32~pc in OMC1 and G327. These sizes are given by the
angular resolution of the current instrumentation at the
hot core distance. Thus, the derived abundances in massive hot cores 
would be severely underestimated in the case of sizes similar
to those of hot corinos. On the other hand, we can have several hot 
cores in the studied region, specially
in massive star forming regions where clustering is more 
important and are located at greater distances. In order to minimize 
these problems, we normalize the studied abundances
to those of the parent species.

As a first step, we have normalized  all the molecular
abundances to that of H$_2$CO and CH$_3$OH. 
These species are thought to be
the $``$parent" molecules of complex O-bearing species,
although some doubts are cast in the particular case of
methyl formate \citep{hor04}. 
In Fig.~7a, we show the relative
abundances of HCOOH, CH$_3$OH, CH$_3$CN and C$_2$H$_5$CN
respect to H$_2$CO as a function of the luminosity. 
Since we have not interferometric observations of H$_2$CO
in FIRS~2,  we have estimated the H$_2$CO abundance
assuming [D$_2$CO]/[H$_2$CO]=0.014. 
A clear trend is observed in the relative abundance of CH$_3$OH, CH$_3$CN and
C$_2$H$_5$CN respect to H$_2$CO as a function of the stellar luminosity
(Fig.~7a). The abundance of all these molecules increases by a factor of $\sim$10
from NGC~1333--IRAS~4A (L=14~L$_\odot$) to OMC1 (L=10$^4$~L$_\odot$).
This trend does not present significant differences between O- and N-bearing
molecules. Contrary to these molecules, the [HCOOH]/[H$_2$CO] ratio 
does not present a systematic behavior.

In Fig.~7b we  plot the abundances of the same molecules normalized to that
of CH$_3$OH. Different behaviors are found for the different molecules.
The [CH$_3$CN]/[CH$_3$OH] ratio does not present
any systematic trend with the protostellar luminosity, although the dispersion in the 
values is quite high. The same remains true for C$_2$H$_5$CN
and probably for C$_2$H$_5$OH (we only have 3 points for C$_2$H$_5$OH). 
This suggests that the ratio between O- and N-bearing complex molecules
does not seem to strongly depend on the protostellar luminosity, contrary to that
expected from theoretical models.
However, the relative abundance of HCOOH
seems to decrease by 2 orders of magnitude from NGC~1333--IRAS4~A to
G327.3. Thus, the relative abundances of HCOOH
to those of CH$_3$OH, CH$_3$CN, C$_2$H$_5$CN and
C$_2$H$_5$OH seem to decrease with the protostellar luminosity.

We propose that the increase of CH$_3$OH/H$_2$CO 
(as well as CH$_3$OH/HCOOH),
CH$_3$CN/H$_2$CO, and C$_2$H$_5$CN/H$_2$CO with 
stellar luminosity is likely due to
differences in the grain mantle composition between low and massive
star forming regions.  Indeed, differences in the initial conditions
(in particular gas density and dust temperature) are known to
strongly affect surface chemistry.  For example, Caselli et al. (1993)
found that complex N-bearing species are more easily formed if the
dust grain temperature (T$_{dust}$) is about 40~K, during the collapse phase.
Although such temperature can be found in a significant fraction
of a collapsing massive clump surrounding a massive young stellar
object \citep{fon02}, this is not the case for low mass
cores.  Therefore, the observed trend (at least for N-bearing species)
is consistent with theoretical predictions. But Caselli et al.
(1993) also found that at T$_{dust}$$\sim$40~K, methanol is not efficiently
formed on the surface because it requires the volatile H to stick
on the grain surface (see also Charnley et al. 1992; van der Tak
et al. 2000).  Thus, the observed CH$_3$OH/H$_2$CO increase
with stellar luminosity is not well understood, unless another
surface formation mechanism for methanol (besides the CO hydrogenation)
becomes important in the warmer high mass clumps, e.g. OH+CH$_3$
(see also discussion in Pontoppidan et al. 2003).  
However, this conclusion is only based on 4 sources and has to be confirmed with
a larger and more complete sample. Observational bias such as
the different spatial scale for the different objects can contribute to
give this trend if the spatial extension of the emission of the 
observed molecules is different.

In Fig.~7c, we compare only the N-bearing molecules.
The [C$_2$H$_5$CN]/ [CH$_3$CN] ratio remains quite constant
with a dispersion of about a factor $<$10 between all the sources.
This uniform  
[C$_2$H$_5$CN]/ [CH$_3$CN] ratio suggesting that the
chemistry of both compounds is linked. Indeed, these two molecules
have a common parent species: C$_2$N (see e.g. Fig.7 in Caselli et al. 1993)

\section{Conclusions: The chemistry of the IM hot core NGC~7129--FIRS~2}
In this paper we present the first detection of an intermediate-mass
hot core.  Our interferometric observations towards FIRS~2
allow us to estimate a size of 650$\times$900~AU and a mass
of 2~M$_\odot$ for the hot core associated with this IM object.
The dimensions  and mass of this IM hot core are intermediate between  
those measured in hot corinos
(r$\sim$150~AU,M$<$1~M$_\odot$) and massive stars 
(r$\sim$3000~AU,M$>$10~M$_\odot$) and consequently one expects
to find a differentiated chemistry in it.

Our interferometric observations of CH$_3$CN, CH$_3$OH,
D$_2$CO and N$_2$D$^+$ provides information on the structure and
chemistry of this object. While the CH$_3$CN emission arises mainly
in the hot core, the CH$_3$OH emission has a component associated with
the bipolar outflow. The CH$_3$CN abundance is $\sim$ 7.0 10$^{-9}$ in the
hot core , which is about 3 orders of
magnitude larger than in the cool envelope.
The methanol abundance is enhanced ($>$3 10$^{-8}$--a few 10$^{-7}$) in 
the hot core and the outflow components. However the rotation 
temperature is higher in the hot core than in the outflow revealing very
different physical conditions, and probably a different CH$_3$OH 
desorption mechanism for the two components. 

The doubly deuterated formaldehyde also presents enhanced
abundances in the hot core. We estimate a [D$_2$CO]/[H$_2$CO]$\sim$0.14, 
which is 4 orders of magnitude larger than the cosmic D
abundance and similar to those found in pre-stellar clumps and
low-mass protostars. The enhanced [D$_2$CO]/[H$_2$CO] abundance in the hot
core suggests that grain surface chemistry is responsible of the deuteration process
of H$_2$CO. In contrast to the high deuteration degree of formaldehyde, we have not 
detected N$_2$D$^+$ in the hot core. This is consistent with the  chemical scheme
in which while the deuteration of H$_2$CO requires surface chemistry,
the deuteration of N$_2$H$^+$ is a gas-phase process.

A large number of molecular lines have been detected in our interferometric spectra
towards FIRS~2. Most of these lines are identified as belonging to deuterated
(D$_2$CO,c-C$_3$D and c-C$_3$HD), sulphuretted ($^{13}$CS,OCS) and complex
O-/N-bearing species (HCOOH,C$_2$H$_5$OH,C$_2$H$_5$CN). Deuterated
species whose deuteration requires surface chemistry, such as D$_2$CO, 
present enhanced 
abundances in the warm regions associated with low-mass 
protostars.  Loinard et al. (2003) searched for the doubly deuterated form of 
formaldehyde (D$_2$CO) in a large 
sample of young stellar objects. D$_2$CO was detected in all low-mass protostars, 
with [D$_2$CO]/[H$_2$CO] 
ratios of 2--40\%. On the other hand, no detection was obtained 
towards more massive protostars (where [D$_2$CO] [H$_2$CO]$<$0.5\%).
This is consistent with the results reported by Turner (1990) who
detected D$_2$CO in Orion and measured [D$_2$CO]/[H$_2$CO]$\sim$0.003.
If the hot cores associated with massive stars are older or
significantly denser than those surrounding low-mass objects, gas
phase chemistry could have had the time to re-set the deuterium
fractionation to values close to the cosmic D/H ratio. An alternative
explanation is that the gas temperature of the material
accreting high mass protostars is larger than 30 K \citep{fon02}, 
so that the deuterium fractionation efficiency in
the gas phase is strongly reduced already before the hot core phase
 
The sulphuretted and complex compounds
are characteristic of hot cores in both the low-mass and the high-mass regimes. 
FIRS~2 is the first IM hot core detected thus far and presents a unique opportunity
to study the link between the chemistry of hot corinos and massive hot cores. 
We have compared the abundances of complex molecules in FIRS~2 with those
in hot corinos and the massive hot cores OMC1 and G327.3--0.6. Contrary to
model predictions, we do no detect any dependence of the O-/N-complex molecules
ratio on the protostellar luminosity. However, we detect differences between the
behavior of the O-bearing species with the stellar luminosity. While H$_2$CO and
HCOOH are more abundant in low luminosity sources, CH$_3$OH 
seems to be more abundant in massive objects. 
We propose that this could be due to a different mantle
composition in the two kinds of region, caused by different
physical conditions (mainly gas density and dust temperature)
during the pre-stellar and accretion phase.
However, this could also be
due to other factors such as the different spatial scale of the observations or
a possible contribution of the shocked gas associated with the bipolar outflow 
to the emission of these molecules. Differences in the hot core ages and/or
cloud initial conditions could also produce these differences. The detection
and detailed study of more
intermediate-mass hot cores are necessary to establish firm conclusions.

\setlength\unitlength{1cm}
\begin{figure}
\vspace{12.5cm}
\includegraphics{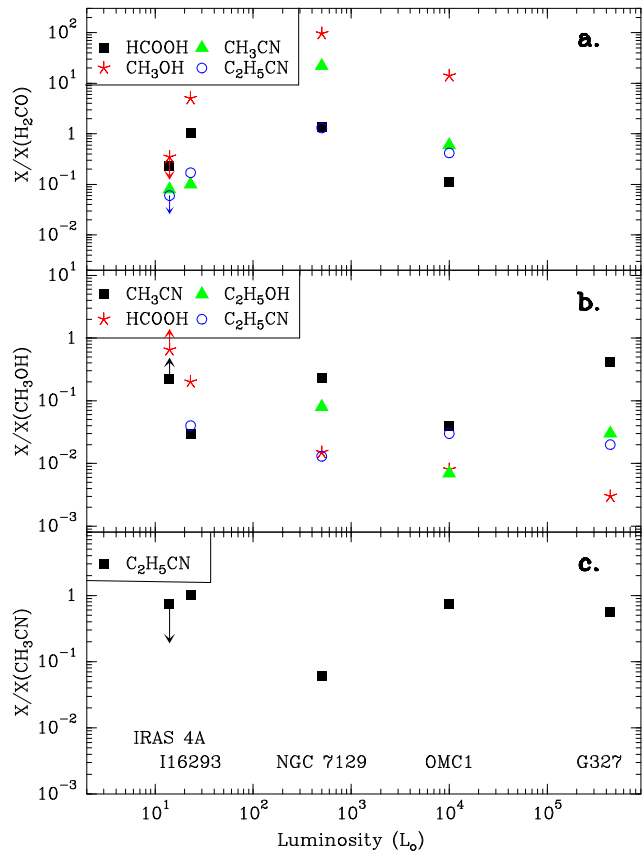}
\caption{Relative abundances of the complex O- and N-bearing molecules
as a function of the protostellar luminosity for the objects listed in Table~3.}
\end{figure}

\acknowledgements
This work has been partially supported
by the Spanish DGICYT under grant AYA2003-07584 and Spanish DGI/SEPCT under
grant ESP2003-04957. PC acknowledges support from the MIUR grant $``$Dust and molecules
in astrophysical environments". We are also grateful to the anonymous referee for his/her
fruitful comments.

\bibliographystyle{aa}
{}
\end{document}